\documentclass[prl,showpacs,twocolumn,amsmath,amssymb]{revtex4}

\usepackage{graphicx}

\newcommand{\dt}{\Delta\tau}

\newcommand{\reff}[1]{Fig.\ \ref{fig:#1}}

\begin{document}

\title{Multigrid Hirsch-Fye quantum Monte Carlo method for dynamical mean-field theory}

\author{N.~Bl\"umer}
\email{Nils.Bluemer@uni-mainz.de}
\affiliation{Institute of Physics, Johannes Gutenberg University, 55099 Mainz, Germany}

\date{\today}

  \begin{abstract}
    We present a new algorithm which allows for direct numerically
    exact solutions within dynamical mean-field theory (DMFT). It is
    based on the established Hirsch-Fye quantum Monte Carlo (HF-QMC)
    method. However, the DMFT impurity model is solved not at fixed
    imaginary-time discretization $\dt$, but for a range of
    discretization grids; by extrapolation, unbiased Green functions
    are obtained in each DMFT iteration. In contrast to conventional
    HF-QMC, the multigrid algorithm converges to the exact DMFT fixed
    points. It extends the useful range of $\dt$, is precise and
    reliable even in the immediate vicinity of phase transitions and
    is more efficient, also in comparison to continuous-time methods.
    Using this algorithm, we show that the spectral weight transfer at
    the Mott transition has been overestimated in a recent density
    matrix renormalization group study.

  \end{abstract}
  \maketitle

  
  Mott metal-insulator transitions and other effects of strong
  electronic correlations are among the most intriguing phenomena in
  solid state physics \cite{SCES}. They occur when the effective
  electronic bandwidths of $d$ or $f$ shell electrons become comparable with
  the local Coulomb interactions. This most interesting regime is also
  most challenging: here, perturbative expansions (both at weak and at
  strong coupling) and effective independent-electron methods such as
  density functional theory \cite{DFT} within local density
  approximation (LDA) break down.

  Unfortunately, nonperturbative methods for a direct treatment of
  correlated lattice electrons are either restricted to
  one-dimensional (i.e., chain- or ladder-like) systems or suffer from
  severe finite-size and/or sign problems. However, a significant
  reduction of complexity in higher-dimensional cases is achieved by
  the dynamical mean-field theory (DMFT) which maps the electronic
  lattice problem onto an Anderson impurity model with a
  self-consistent bath; this mapping becomes exact in the limit of
  infinite lattice coordination \cite{Georges96}. Within the last 15
  years, much insight into strongly correlated systems and phenomena
  has been gained using the DMFT. Many of these studies have relied on
  the Hirsch-Fye quantum Monte Carlo (HF-QMC) method \cite{Hirsch86}
  for solving the DMFT impurity problem at finite temperatures. 

  The HF-QMC method discretizes the imaginary-time path integral into
  slices of uniform width $\dt$ and employs a Trotter decoupling; this
  modified impurity problem is solved via Monte Carlo sampling of a
  binary Hubbard-Stratonovich field. In principle, arbitrary precision
  can be achieved using HF-QMC in the combined limit of infinitely
  many updates of the Hubbard-Stratonovich field and of vanishing
  discretization $\dt$; in this broader sense, HF-QMC is {\em
    numerically exact}. However, practical lower limits for the
  discretization exist (primarily due to a scaling of the numerical
  effort with $(\dt)^{-3}$ for fixed temperature $T$), so that raw
  HF-QMC results usually contain systematic Trotter errors which are
  much larger than the statistical Monte Carlo errors. This implies,
  in the DMFT context, that all observables and phase boundaries
  depend on the auxiliary parameter $\dt$ chosen in the
  self-consistency cycle.  While high accuracy and efficiency have
  been demonstrated for {\em a posteriori} extrapolations $\dt\to 0$
  of certain static observables \cite{Bluemeretc,Bluemer07}, these
  procedures require special care and experience and may fail close to
  phase transitions. Up to recently, the Trotter bias of HF-QMC
  derived spectral functions could not be reduced at all.

  For a long time, Trotter errors could only be avoided using
  fundamentally different impurity solvers that either introduce
  a logarithmic frequency discretization directly (numerical
  renormalization group, NRG) or represent impurity plus bath as a finite cluster
  or chain [exact diagonalization (ED),
  density-matrix renormalization group (DMRG)]. Consequently, their
  results, too, contain systematic finite-size/discretization errors;
  in particular, these alternatives to QMC yield continuous spectra
  only after numerical broadening. Only very recently, two new quantum
  Monte Carlo impurity solvers have become available
  \cite{Rubtsov05,Werner06} which are {\em numerically exact} in the
  stricter sense that their results are correct within statistical
  error bars, without the need for explicit extrapolations. These
  methods avoid the imaginary-time discretization of HF-QMC and rely,
  instead, on perturbative expansions in the interaction and
  hybridization, respectively, which are statistically sampled to
  arbitrary order.  However, the continuous-time quantum Monte Carlo
  (CT-QMC) methods are often less efficient than HF-QMC
  \cite{Bluemer07}.

  In this Letter, we propose a new algorithm for solving the DMFT
  self-consistency equations in quasi continuous imaginary time, based
  on a multigrid implementation of the HF-QMC method. This
  implementation removes many of the limitations of HF-QMC; in
  particular, it yields precise and reliable results even at phase
  boundaries. As a first application, we test DMRG predictions \cite{Karski05} of the
  spectral weight transfer at the Mott transition -- a study that
  would not have been possible using conventional HF-QMC.

  
  {\it Multigrid HF-QMC method --} Before we formulate the new
  multigrid approach, let us discuss the flow diagram
  \reff{DMFTcycleMulti}(a) of the conventional HF-QMC method.
  \begin{figure}
    \includegraphics[width=\columnwidth]{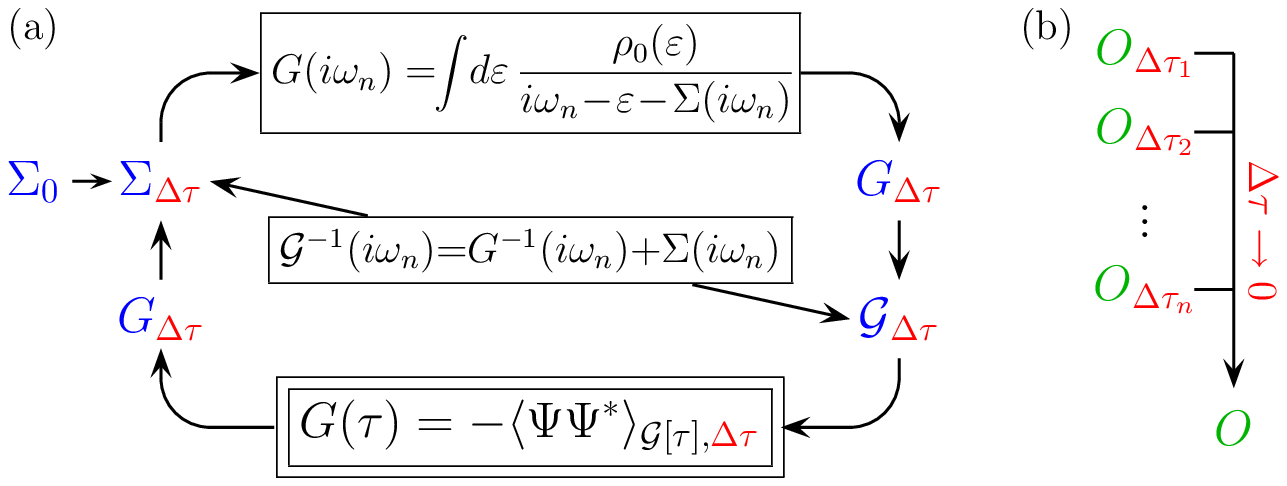}

    \vspace{2.5ex}
    \includegraphics[width=\columnwidth]{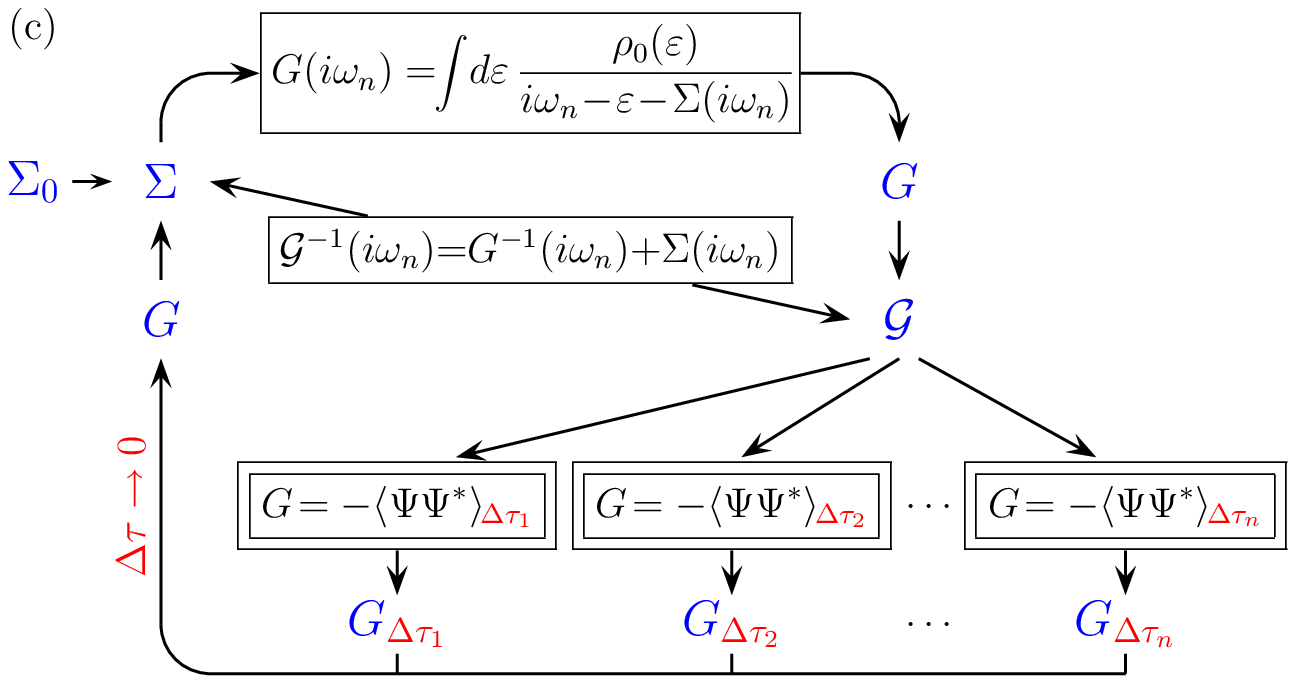}
    \caption{(Color online) 
      (a)~Conventional HF-QMC algorithm: Trotter error of impurity
      Green function $G$ (computed at fixed $\dt$) leads to bias in
      bath Green function ${\cal G}$, self-energy $\Sigma$, and all
      observables. (b)~Extrapolation $\dt\to 0$ of observable {\cal
        O}. (c) Multigrid HF-QMC scheme: DMFT iteration with unbiased
      impurity Green function, extrapolated from multiple HF-QMC
      solutions $G_{\dt_1} \dots G_{\dt_n}$: at self-consistency, $G$,
      ${\cal G}$, $\Sigma$, and all derived observables are numerically
      exact.  }\label{fig:DMFTcycleMulti}
  \end{figure}
  Starting with some guess $\Sigma_0$ for the self-energy, the lattice
  Green function $G$ is obtained via the lattice Dyson equation (upper
  box, here written for the 1-band Hubbard model); as a next step, the
  impurity Dyson equation (middle box) is used to determine the bath
  Green function ${\cal G}$ which defines the DMFT impurity problem
  (lower box).  Conventionally, this is solved using HF-QMC at a fixed
  discretization; the resulting Green function closes the DMFT cycle
  via the impurity Dyson equation. Obviously, the Trotter error in the
  impurity solver affects the whole DMFT cycle; thus, at
  self-consistency, $G_{\dt}$, ${\cal G}_{\dt}$, $\Sigma_{\dt}$ and
  all observables deviate from their physical values, ideally with
  corrections in powers of $(\dt)^2$. Only then, numerically exact
  estimates of observables can be extrapolated {\em a posteriori}
  from results of independent HF-QMC DMFT runs, as
  illustrated in \reff{DMFTcycleMulti}(b).

  The new multigrid method, visualized in \reff{DMFTcycleMulti}(c),
  splits the impurity-solving part of the DMFT self-consistency cycle
  into two steps: first, the impurity problem is solved -- at fixed
  bath Green function ${\cal G}$ -- for a suitable range of
  discretizations $\dt_i \in [\dt_{\text{min}},\dt_{\text{max}}]$
  using HF-QMC. In a second step, a numerically exact estimate $G$ of
  the true Green function of the given impurity problem is obtained by
  extrapolation of the multiple resulting Green functions $G_{\dt_i}$.
  This highly nontrivial task is accomplished using a scheme developed
  recently (in the context of {\em a posteriori} extrapolation)
  \cite{Bluemer07a}: (i) each of the discrete HF-QMC estimates is
  interpolated with the help of a suitably chosen reference model onto
  a common fine $\tau$ grid; (ii) quadratic (in $\dt^2$) least-squares
  extrapolations $\dt\to 0$ are performed on this grid. In practice,
  the raw Green functions $G_{\dt_i}$ are averaged over 10--20
  impurity solutions.  Thus, for about 10 discretizations $\dt_i$, the
  multigrid method parallelizes to about 100 CPU cores.  For optimal
  accuracy, a hierarchy of discretization scales and high-frequency
  cut-offs needs to be maintained \cite{fn:parameters}; details will
  be presented elsewhere \cite{BluemerPrep}. Since the multigrid
  algorithm closes the DMFT self-consistently with an impurity Green
  function that is (for correctly chosen parameters, see below)
  unbiased, all state variables $G$, ${\cal G}$, and $\Sigma$ converge
  to their numerically exact forms.

  This fundamental advantage of the multigrid HF-QMC method is illustrated
  in the scheme \reff{FreeMulti}:
  \begin{figure}
    \includegraphics[width=\columnwidth]{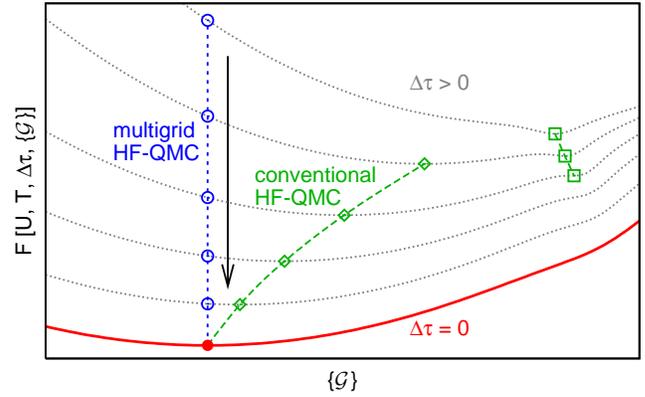}
    \caption{(Color online) Scheme: GL free energy in multidimensional
      space of hybridization functions $\{{\cal G}\}$. The fixed
      points of conventional HF-QMC (diamonds, squares) are
      adiabatically connected to the exact fixed point (full circle)
      only for small $\dt$. In contrast, the multigrid method solves
      all impurity problems at the $\dt=0$ fixed point
      (circles).}\label{fig:FreeMulti}
  \end{figure}
  Here, the solid line depicts the exact Ginzburg-Landau (GL) free
  energy $F[U,T,\{{\cal G}\}]$ in the multi-dimensional space of bath
  Green functions $\{{\cal G}\}$ (for fixed physical parameters, $U$,
  $T$); its stationary points mark the DMFT fixed points
  \cite{Kotliar00}. However, conventional HF-QMC implementations
  converge to the stationary points (squares, diamonds) of generalized
  GL functionals (dotted lines) with additional dependence on $\dt$.
  These fixed points can vary irregularly or even discontinuously as a
  function of $\dt$ which complicates or prevents extrapolations
  $\dt\to 0$.  In contrast, the multigrid HF-QMC procedure drives the
  DMFT iteration to the numerically exact fixed point: at
  self-consistency, each finite-$\dt$ HF-QMC evaluation is performed
  for the bath Green function at which $F[U,T,\{{\cal G}\}]$ is
  stationary (full circle) while the generalized functionals
  $F[U,T,\dt, \{{\cal G}\}]$ are not (empty circles).

\begin{figure}
    \includegraphics[width=\columnwidth]{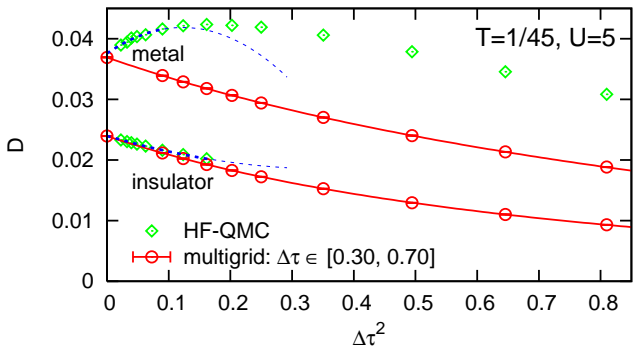}
    \caption{(Color online) Double occupancy $D=\langle n_\downarrow
      n_\uparrow\rangle$ at $T=1/45$, $U=5$ for metallic (upper set of
      curves) and insulating (lower set of curves) phases: results
      from conventional HF-QMC (diamonds) and from multigrid HF-QMC
      (circles).}\label{fig:DoubleB45U50}
  \end{figure}

  
  {\it Benchmark results --} For a quantitative discussion, let us now
  consider the one-band Hubbard model with semi-elliptic density of
  states (bandwidth $W=4$) at low temperatures $T$ in the
  strongly correlated regime. Specifically, we will concentrate on the
  double occupancy $D=\langle n_\downarrow n_\uparrow\rangle$, an
  observable which is well-defined for the impurity model irrespective
  of DMFT self-consistency 
  and which is best computed at fixed $\dt$ in both algorithms.
  Figure \ref{fig:DoubleB45U50}
  shows results for the coexisting metallic and insulating phases
  versus squared discretization at $T=1/45$ and interaction $U=5$.
  Conventional HF-QMC (diamonds) finds an insulating solution only for
  relatively small $\dt\lesssim 0.4$. While a metallic solution can be
  stabilized at arbitrarily large $\dt$, the $\dt$ dependence of $D$
  is highly nonuniform which restricts the useful range for
  extrapolations $\dt\to 0$ (dashed lines), again, to small values
  $\dt\lesssim 0.4$.  In contrast, the multigrid HF-QMC raw data
  (circles) shows regular $\dt$ dependence even at large
  discretizations which allows for very precise extra\-polations (solid
  lines) based on cheap large-$\dt$ HF-QMC data \cite{fn:logD}. Evidently,
  the multigrid method is superior.

  However, such extrapolations, as well as all observable estimates
  that can be derived from the Green function and self-energy without
  explicit $\dt\to 0$ extrapolation, are only reliable if the
  multigrid algorithm has really converged to the exact DMFT fixed
  point. Thus, we have to check for any bias that might survive from
  the intrinsic parameters, most notably the range
  $[\dt_{\text{min}},\dt_{\text{max}}]$ used in the internal Green
  function extrapolation. In the inset of
  \reff{DoubleB45U40m},
  \begin{figure}
    \includegraphics[width=\columnwidth]{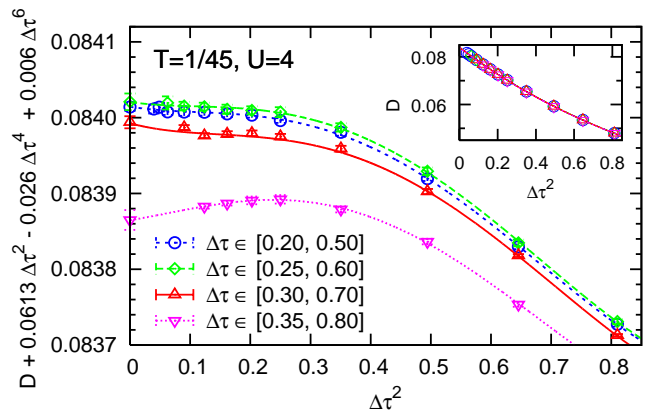}
    \caption{(Color online) Inset: multigrid HF-QMC results for double
      occupancy $D=\langle n_\downarrow n_\uparrow\rangle$ at
      $T=1/45$, $U=4$, using different ranges $[\tau_{\text{min}},
      \tau_{\text{max}}]$ in the Green function extrapolation. Main
      panel: same data minus leading $\dt$
      corrections.}\label{fig:DoubleB45U40m}
  \end{figure}
  multigrid HF-QMC results (at $U=4$, i.e, in the metallic phase) are
  plotted for various $\dt$ grids; at this scale, both the raw data
  and the fit curves fall on top of each other.  For a fine-scale
  analysis, the approximate leading Trotter errors have been
  subtracted in the main panel of \reff{DoubleB45U40m}. Even for this
  extremely precise raw data (with $2\cdot 10^{7}$ sweeps for each
  $\dt$ in 10 iterations after convergency), no impact of the grid
  range can be detected for $\tau_{\text{min}}\le 0.25$.  A slight
  bias of the order $10^{-5}$ is visible for the grid range
  $[0.3,0.7]$; even the coarsest range $[0.35,0.8]$ of discretizations
  shifts $D$ only by about $10^{-4}$.  Similar studies indicate even
  slightly smaller grid range effects in the insulating phase (at
  $U=5$) \cite{BluemerPrep}. Thus, the
  multigrid HF-QMC method can be considered numerically exact for grid
  ranges up to $[0.3,0.7]$ in almost all cases; finer grids or a
  detailed analysis are only needed when extreme precision is sought.
  This should be contrasted with the conventional HF-QMC method for
  which biases in Green functions and self-energies should remain
  detectable down to $\dt\approx 0.01$ -- if reasonable statistical accuracy
  could be maintained.


  {\em Spectral weight transfer at the Mott transition --} While many
  aspects of the Mott metal-insulator transition are well-established
  by now, including the phase diagram of the half-filled frustrated
  one-band model within DMFT, some fundamental questions have remained
  open.  In particular, it is still not clear how exactly the spectra
  change across the first-order Mott transition, where the quasiparticle
  peak disappears. Recently, a new scenario has been suggested on
  the basis of dynamical density matrix renormalization group (DDMRG)
  calculations \cite{Karski05}. As seen in the inset of
  \reff{dosU52},
  \begin{figure}
    \includegraphics[width=\columnwidth]{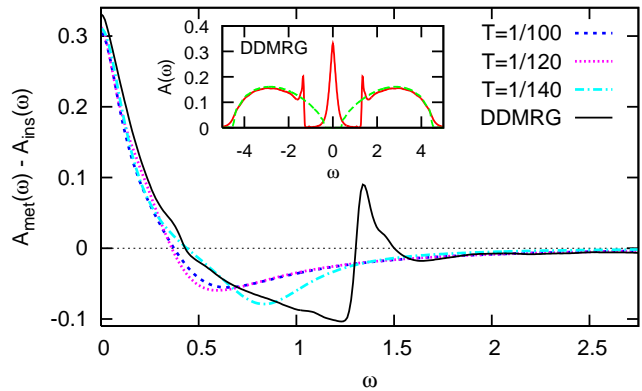}
    \caption{(Color online) Inset: DDMRG spectra of coexisting
      metallic and insulating phases at $U=5.2$, reproduced from
      \cite{Karski05}. Main panel: difference spectra from multigrid HF-QMC
      (broken lines) and DDMRG \cite{Karski05} (solid line).}\label{fig:dosU52}
  \end{figure}
  this study finds sharp features at the inner edges of the Hubbard
  bands in the strongly correlated metallic phase (solid lines), which
  are separated from the quasiparticle peak by quite broad pseudo
  gaps; in contrast, the Hubbard bands appear rather featureless in
  the insulating phase (dashed lines) for the same interaction
  $U=5.2$, with inner edges that are shifted to much smaller
  frequencies.  Obviously this behavior, which has been interpreted in
  terms of collective magnetic excitations in the metal
  \cite{Karski05}, should be verified. One may also ask whether it
  extends to finite temperatures, i.e., governs the spectral weight
  transfer at the Mott transition, which, for $U=5.2$, occurs at
  $T\approx 1/100$.

  Estimates of the spectral weight transfer, the difference of
  metallic and insulating spectra, are shown in the main panel of
  \reff{dosU52}. At this scale, the DDMRG data (solid line) clearly
  resolves the features discussed above, in particular the Hubbard
  subpeak at $|\omega|\approx 1.3$. The corresponding multigrid HF-QMC
  results (broken lines) have been obtained, for minimal bias, via
  Pad{\'e} analytic continuation of Matsubara difference Green
  functions.  They are much smoother even at very low $T$: apart from
  the quasiparticle peak, with a reduced weight, only a shallow
  negative region at $0.4\lesssim \omega \lesssim 1.5$ is visible,
  which is evidence against the existence of subpeaks. We obviously cannot
  exclude that this smoothness is, at least partially, an artifact of
  the ill-conditioned analytic continuation; it is also not clear whether
  the change of shape at the lowest temperature $T=1/140$, with a
  closer approach of the DDMRG line shape, is significant.
  In contrast, the QMC difference Green functions shown in \reff{GtauU52}
  as broken lines (basis for the spectral data of \reff{dosU52}) are
  precise within linewidth. In the low-$\tau$ regime that we are
  focusing on, even the extrapolation to $T=0$ is reliable (see inset
  of \reff{GtauU52}). However, the transformed DDMRG data (thin solid
  line in main panel) deviates markedly from this QMC ground state
  result (thick solid line, with a precision of about linewidth): the
  DDMRG overestimates the differences between metal and insulator, in
  this measure, by about $10\%$, with an even larger discrepancy
  compared to the finite-temperature result at $T=1/100$, where the
  metal-insulator transition takes place. We conclude that the DDMRG
  does not capture the spectral-weight transfer at the Mott transition
  {\em quantitatively} correctly; possible explanations include the
  bias towards metallic/insulating solutions for odd/even chains in
  the DMRG calculation and, in particular, the deconvolution
  technique. In terms of spectra, a large part of the DDMRG error can
  be traced back to its overestimation of the quasiparticle peak
  (since the QMC data of \reff{dosU52} is particularly reliable in
  this range); similar mechanisms might have generated the
  Hubbard-band subpeaks.
  \begin{figure}
    \includegraphics[width=\columnwidth]{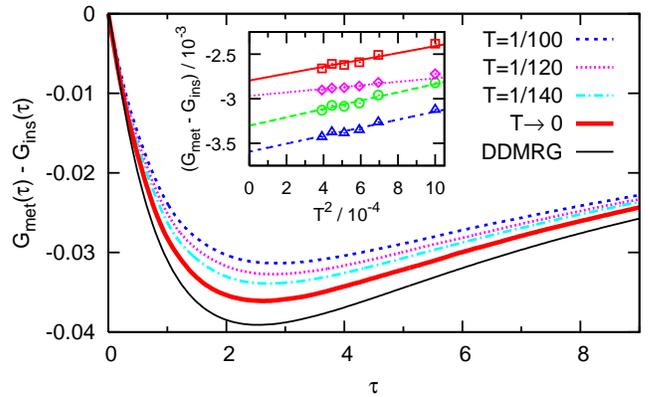}
    \caption{(Color online) Difference Green functions from QMC at
      finite $T$ (dashed, dotted lines) and extrapolated to $T=0$
      (thick solid line) versus DDMRG data (thin line). Inset:
      extrapolation $T\to 0$ for $\tau=1, 6, 1.5, 2.5$ (top to
      bottom).}\label{fig:GtauU52}
  \end{figure}


  {\em Discussion --} We have formulated a new algorithm for solving
  Hubbard-type models within DMFT, based on a multigrid implementation
  of the HF-QMC method with internal Green function extrapolation.
  This technique directly yields numerically exact results even at
  very low temperatures, similarly to recently developed
  continuous-time QMC methods. However, our algorithm retains the
  advantages of HF-QMC, i.e., generates Green functions with minimal
  fluctuations, now at about halved temperatures for similar precision
  and effort. The method extends to the general multi-band case, with
  enormous potential, e.g., in the context of {\em ab initio} LDA+DMFT
  calculations. While it is too early to judge which of the
  three (quasi) continuous-time methods will ultimately prevail, our
  multigrid method clearly supersedes conventional HF-QMC for most
  applications.

  Using this method, we have found inconsistencies in the DDMRG
  scenario of the spectral weight transfer at the Mott transition.  A
  definite judgement whether the DDMRG results are, at least, {\em
    qualitatively} correct, will require additional computing
  ressources and detailed analyses of analytical continuation procedures.

  Stimulating discussions with P.G.J.\ van Dongen and support by the
  DFG within the Collaborative Research Centre SFB/TR 49 are
  gratefully acknowledged.


\begin{thebibliography}{16}
\bibitem{fn:logD} We have verified that the assumed extrapolation law $\log[D(\dt)]=\sum_{n=0}^N d_n \dt^{2n}$ remains accurate at least up to $\dt=2$,
when $N$ is increased from 3 to about 8.
\end{thebibliography}
\end{document}